\documentclass[letterpaper]{article} 
\usepackage[submission]{aaai25}  
\usepackage{times}  
\usepackage{helvet}  
\usepackage{courier}  
\usepackage[hyphens]{url}  
\usepackage{graphicx} 
\usepackage{natbib}
\usepackage{caption}
\usepackage{hyperref}
\usepackage{algorithm}
\usepackage{algorithmic}

\usepackage{newfloat}
\usepackage{listings}
\DeclareCaptionStyle{ruled}{labelfont=normalfont,labelsep=colon,strut=off} 
\floatstyle{ruled}
\newfloat{listing}{tb}{lst}{}
\floatname{listing}{Listing}

\usepackage{booktabs}
\usepackage{multirow}

\title{Poplar: Efficient Scaling of Distributed DNN Training on Heterogeneous GPU Clusters}
\author {
    Wenzheng Zhang\textsuperscript{\rm 1},
    Yang Hu\textsuperscript{\rm 2},
    Jing Shi\textsuperscript{\rm 1},
    Xiaoying Bai\textsuperscript{\rm 3}
}

\affiliations {
    \textsuperscript{\rm 1}School of Computer Science, Peking University\\
    \textsuperscript{\rm 2}Information Research Center of Military Science, PLA Academy of Military Science\\
    \textsuperscript{\rm 3}Advanced Institute of Big Data\\
    zwz@stu.pku.edu.cn, huyang.oceans@gmail.com, jingshi@stu.pku.edu.cn, baixy@aibd.ac.cn
}

\usepackage{bibentry}

\begin{document}

\maketitle

\begin{abstract}
Scaling Deep Neural Networks (DNNs) requires significant computational resources in terms of GPU quantity and compute capacity. In practice, there usually exists a large number of heterogeneous GPU devices due to the rapid release cycle of GPU products. It is highly needed to efficiently and economically harness the power of heterogeneous GPUs, so that it can meet the requirements of DNN research and development. The paper introduces Poplar, a distributed training system that extends Zero Redundancy Optimizer (ZeRO) with heterogeneous-aware capabilities. We explore a broader spectrum of GPU heterogeneity, including compute capability, memory capacity, quantity and a combination of them. In order to achieve high computational efficiency across all heterogeneous conditions, Poplar conducts fine-grained measurements of GPUs in each ZeRO stage. We propose a novel batch allocation method and a search algorithm to optimize the utilization of heterogeneous GPUs clusters. Furthermore, Poplar implements fully automated parallelism, eliminating the need for deploying heterogeneous hardware and finding suitable batch size. Extensive experiments on three heterogeneous clusters, comprising six different types of GPUs, demonstrate that Poplar achieves a training throughput improvement of 1.02 $\sim$ 3.92x over current state-of-the-art heterogeneous training systems.
\end{abstract}

\section{Introduction}
With the increase in model parameters, the memory and compute requirements for model training have grown significantly beyond the capability of a single accelerator\cite{wang2023zero++}. Training large-scale models requires the efficient utilization of aggregated computing power and memory across hundreds or even thousands of GPUs\cite{smith2022using}. There are two distributed training frameworks to this, namely 3D Parallelism\cite{shoeybi2019megatron, lai2023merak} and Zero Redundancy Optimizer(ZeRO)\cite{rajbhandari2020zero}. 3D Parallelism requires a complex and meticulous design by domain experts\cite{zheng2022alpa}. Conversely, ZeRO needs no code refactoring, which has contributed to its widespread adoption in model training\cite{smith2022using}.

\begin{figure}[t]
\centering
\includegraphics[width=1\columnwidth]{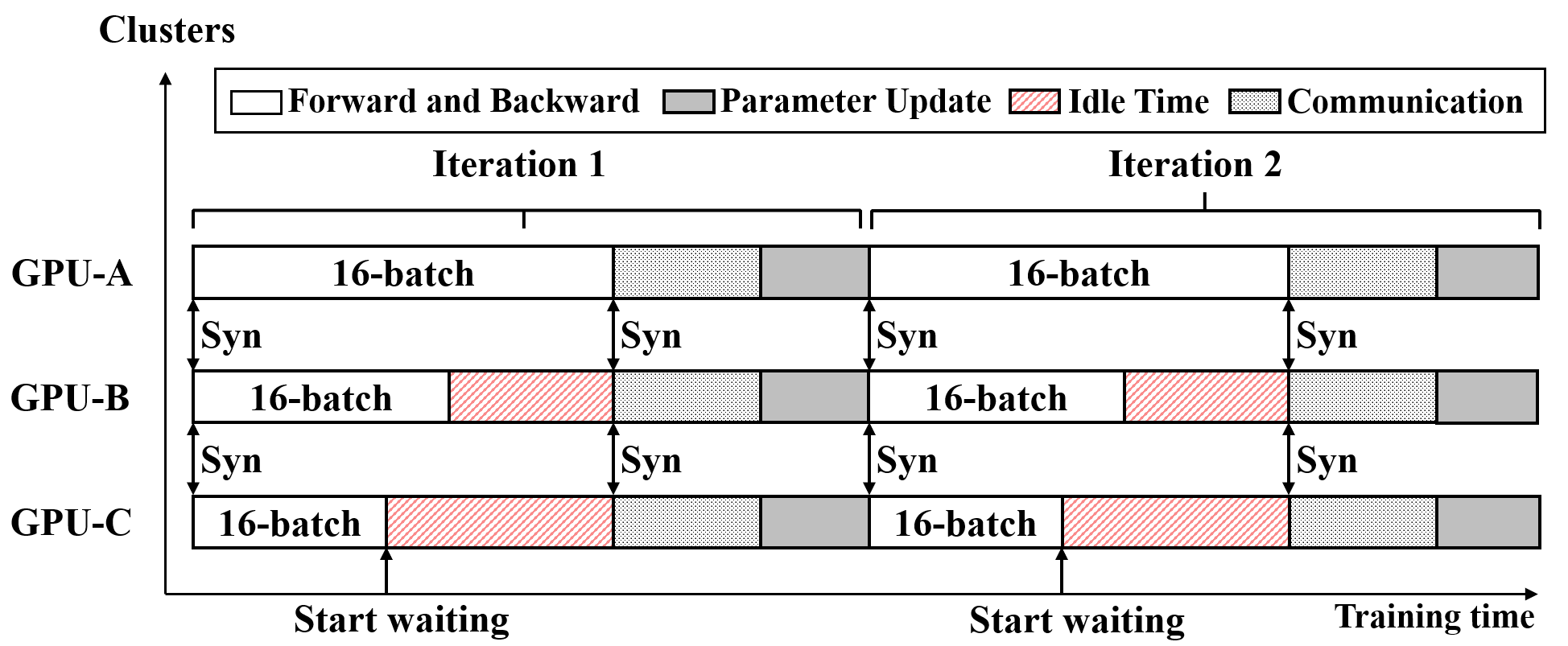} 
\caption{High-end GPUs complete their tasks first and then start waiting before synchronization. Without load balancing, there will be significant idle time.}
\label{straggler}
\end{figure}

Acquiring a large number of homogeneous high-end GPUs within a cluster often requires a long queuing time, while a sufficient quantity of heterogeneous GPUs is more readily available\cite{weng2022mlaas}. Some researchers have access only to a variety of consumer-grade GPUs that cannot individually support the training of large models\cite{song2023powerinfer}. Furthermore, the cost of purchasing new machines is substantial due to the rapid release cycle of GPU products\cite{choquette2022nvidia}. In such situations, effectively utilizing heterogeneous GPU resources can help address these challenges, thereby speed up the model exploration and experiments. Current techniques and systems for distributed model training mostly assume that the workers are homogeneous\cite{shoeybi2019megatron, rasley2020deepspeed}. As Figure \ref{straggler} showed, directly applying these techniques in heterogeneous clusters will result in substantial idle time during synchronization\cite{kwon2020nimble}, since powerful GPUs must wait for less ones. Achieving load balancing to minimize the idle time of each GPU in heterogeneous clusters is challenging\cite{tang2021aeml}, since both the compute capability and the memory capacity of GPUs need to be considered. 

Many studies have focused on this issue, AMP\cite{li2022amp} and Whale\cite{jia2022whale} incorporate the elements of heterogeneity into search space of auto-parallel algorithms. AMP is equipped with an expert-design cost model that considers cluster and model configurations. Whale uses automatic graph optimizations to adapt heterogeneous GPUs\cite{jia2019beyond}. Whale and AMP can enhance the utilization of heterogeneous clusters based on 3D parallelism.

However, previous research only addresses limited aspects of heterogeneity. They only performs well on heterogeneous GPUs with varying compute capabilities and memory capacities(e.g., V100 and one A100). This restricts the effective utilization of actual heterogeneous GPU clusters. Due to the nature of 3D parallelism, existing methods struggle in two scenarios: (1)where only memory capacities differ while compute capability remain equal(e.g., A100-80GB and A100-40GB), and (2) where the number of heterogeneous GPUs is non-uniform(e.g., two V100 and one A100). Additionally, Whale uses single-precision floating point operations per second (FLOPs) as cost model to measure the compute capability of GPUs. However, FLOPs can not accurately reflect performance differences during training between heterogeneous GPUs, potentially introducing a degree of error in performance evaluations. Meanwhile, manually configuring heterogeneous hardware and batch size for training job in heterogeneous GPU clusters is impractical, due to the increased numbers of variables in 3D parallelism compared to homogeneous clusters.

To address these challenges, we propose a distributed training system, named Poplar. Firstly, we extend ZeRO to support heterogeneous GPUs, while considering a wider range of GPU heterogeneity, including compute capability, memory capacity, quantity and their combinations. We treat each GPU as an individual unit and assign tasks separately to ensure maximum global throughput. Secondly, to bridge the gap between cost model and real-world performance, we conducte a fine-grained analysis for each ZeRO stage and propose a novel method for measuring heterogeneous of GPUs. We also develop a search algorithm alone with a batch allocation method to ensure load balancing. Thirdly, we enable automated determination of the optimal configuration across heterogeneous GPUs, eliminating the need for expert experience and manual adjustments. The experiments result on three real world heterogeneous GPU clusters demonstrated that Poplar can speed up the training process 1.02 $\sim$ 3.92x than other methods. We will publish all source codes of this work on Github for further research explorations. In summary, Our contributions are as follows:
\begin{itemize}
    \item To the best of our knowledge, we are the first to investigate training LLM with ZeRO on heterogeneous GPUs. We designed and implemented a distributed DNN training system, Poplar, which extend ZeRO to harness the power of heterogeneous GPUs.
    \item We explore a broader spectrum of GPU heterogeneity, and develop a heterogeneity aware method along with an batch size searching algorithm for each ZeRO stage to improve cluster utilization.
    \item Poplar enables fully automated parallel training, allowing for automatic determining the optimal training configuration in heterogeneous GPU clusters.
    \item We conduct extensive experiments with diverse sets of real world heterogeneous GPU clusters. The results demonstrated that Poplar can speed up the training process 1.02 $\sim$ 3.92x than other methods.
\end{itemize}

\section{Related Work}
\subsection{Zero Redundancy Optimizer}
ZeRO is a variant of data parallelism based on model parallelism, which can substantially reduce the memory required for training models. ZeRO partitions optimizer states among GPUs in ZeRO-1, distributes gradients in ZeRO-2, and places parameter shards on different GPUs in ZeRO-3. ZeRO is a distributed training system based on Bulk Synchronous Parallel\cite{gerbessiotis1994direct}, where all computational units maintain consistency in model parameters after each round of gradient updates. A detailed analysis of ZeRO's process is provided in Appendix.

\begin{figure*}[!t]
\centering
\includegraphics[width=2\columnwidth]{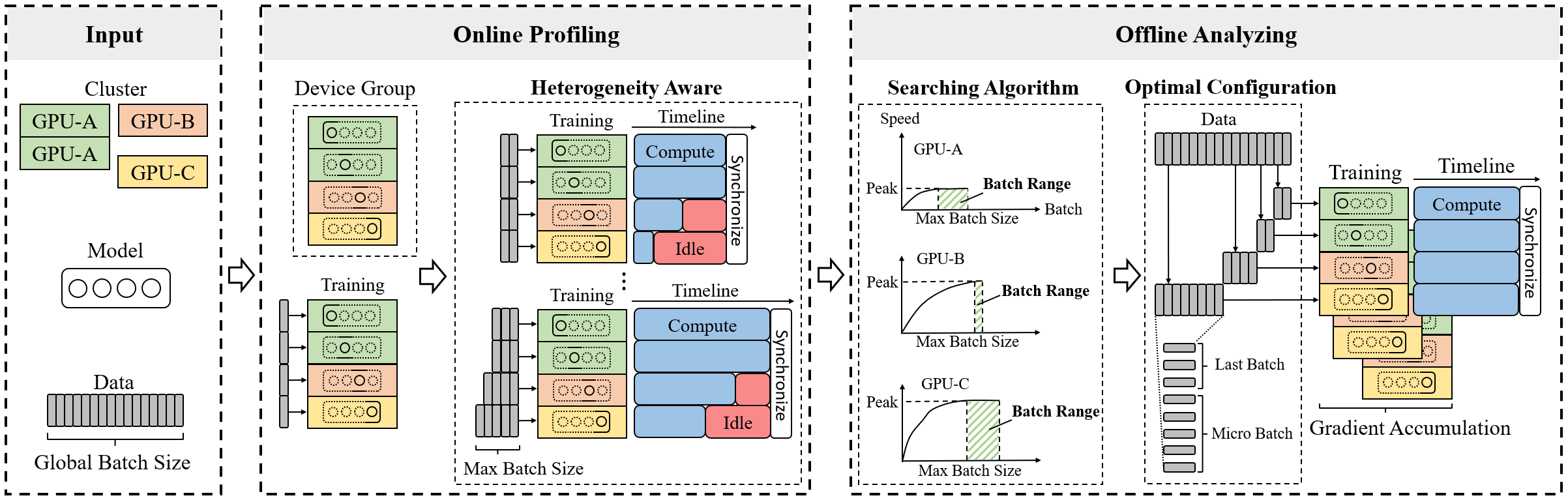}
\caption{An overview of how Poplar automatically determines the optimal configuration..}
\label{SystemOverview}
\end{figure*}

\subsection{Training in Heterogeneous GPU Clusters}

Some researches use asynchronous synchronization\cite{zhou2020petrel} to alleviate the challenges posed by heterogeneity, such as HetPipe\cite{park2020hetpipe}, SAP-SGD\cite{cao2021sap}, Heter-Train\cite{geng2023heter}, LB-BSP\cite{chen2021accelerating}and Hop\cite{luo2019hop}. Nonetheless, the approach of asynchronously updating gradients lacks a rigorous mathematical proof of Convergence. Whale\cite{jia2022whale} introduced a novel hardware-aware parallel strategy to enable auto parallelism. AMP\cite{li2022amp} develop a dynamic programming approach to handle the load imbalance issue from heterogeneous models in the pipeline layer assignment problem. Meanwhile, Whale and AMP impose restrictions on the number of different GPUs, failing to consider scenarios with non-uniform numbers of heterogeneous devices. These limitations greatly constrain the application in real-world heterogeneous GPU cluster.

\section{Method}
Training a well performed model requires a substantial amount of data. The model operates on a portion of the dataset and update parameters in each iteration. The entire training process always consists of many iterations. For example, LLaMA-2 requires 500,000 iterations to train a model with 7B parameters\cite{touvron2023llama}. Within each iteration, the model's operations are repetitive. The goal of Poplar is to minimize the time required for every single iteration, thereby reducing the overall training time and improving the hardware utilization.

\subsection{Problem Formulation}
Formally, the inputs of Poplar are (i) the model, (ii) the GPU cluster, and (iii) the global batch size ($gbs$). During each iteration of the training process, the model needs to process the data of size $gbs$. We consider each GPU in the cluster as an independent device, the cluster can be represented as $ Cluster = \{g_1, g_2, g_3, \cdots, g_n\}$, where $n$ is the totel number of GPUs. Meanwhile, the compute capability of each GPU is denoted as $Speed = \{p_1, p_2, p_3, \cdots, p_n\}$. We need to find a configuration $B$ such that $Cluster$ can process $gbs$ data as quickly as possible. The configuration $B$ can be represented as follows: 
$$ B = \{b_1, b_2, b_3, \cdots, b_n\}, \sum_{i=0}^n b_i = gbs $$ 
$$ 0 \leq b_i \leq mbs_i, i=1,2,\cdots,n $$
$mbs_i$ represents the max batch size of GPU $g_i$ can handle without causing an out-of-memory (OOM) error. Each GPU has a different runtime $t_i$ during a single iteration. Synchronization is required among heterogeneous GPUs in each iteration, which involves waiting until all GPUs have completed their tasks. Consequently, the iteration time $T$ and idle time $\delta t_i$ of each GPU can be calculated as:
\begin{equation}
    T = max\{t_1, t_2, t_3, \cdots, t_n\}
\label{wall time}
\end{equation}
\begin{equation}
    \delta t_i = T - t_i, i=1,2,\cdots,n
\label{delta time}
\end{equation}
To ensure the shortest computing time $T$, our objective is to identify the configuration $B$ that maximizes overall GPU utilization. In other words, we aim to find the configuration that minimizes overall GPU under-utilization. The under-utilization of each GPU is denoted as $u_i$ which can be calculated as:
\begin{equation}
    u_i = \delta t_i \times p_i, i=1,2,\cdots,n
\label{under utilization}
\end{equation}
Powerful GPUs typically own a larger $s$, and therefore have a greater impact on the entire cluster. Then the objective function of Poplar will be:
\begin{equation}
    \mathop{\arg\min}\limits_{b_1, b_2, b_3, \cdots, b_n} \sum_{i=1}^{n} \delta t_i \times p_i, i=1,2,\cdots,n
\label{objentive function}
\end{equation}

\subsection{System Overview}
Figure \ref{SystemOverview} illustrates the overall structure of Poplar. Initially, the model, GPU clusters and global batch size serve as inputs to Poplar. In the online profiling, Poplar detect the heterogeneity among all GPUs precisely, preparing the data for the subsequent offline analyzing phase. Poplar monitors metrics for each GPU during the runtime process, where the memory capacity is quantified by the maximum batch size($mbs$) of GPU, and the compute capability is reflected by the corresponding processing time. Afterwards, during the offline analyzing phase, Poplar use a batch allocation method and a search algorithm to determine the optimal allocation configuration. Based on this configuration, tasks will be assigned to each GPU to start training. The above process is a general workflow for ZeRO to support heterogeneous environment. Different ZeRO stages have distinct process in both two phase, which will be detailed in the following sections.

\begin{algorithm}[!tb]
    \renewcommand{\algorithmicrequire}{\textbf{Input:}}
	\renewcommand{\algorithmicensure}{\textbf{Output:}}
	\caption{Heterogeneity Aware of each GPU}
    \label{power}
    \begin{algorithmic}[1]
        \REQUIRE  GPU cluster: $Cluster = \{g_1, g_2, \cdots, g_n\}, model$
	    \ENSURE compute capability: $\mathbf{P} = \{p_1, p_2, \cdots, p_n\}$,\\ max batch size: $ \mathbf{Mbs} = \{mbs_1, mbs_2, \cdots, mbs_n\}$

        \FOR{each GPU $g_i$ in $Cluster$ \textbf{in parallel}} 
            \STATE $batch\_size_i \gets 1$
            \STATE $bf_i \gets$ CurrentMemoryAlloced()
            \STATE model.forward($batch\_size_i$)
            \STATE $af_i \gets$ CurrentMemoryAlloced()
            \STATE $memory_i \gets$ TotalMemory()
            \STATE $mbs_i \gets (memory_i - bf_i) / \frac{af_i - bf_i}{batch\_size_i}$
        \ENDFOR
        \FOR{each GPU $g_i$ in $Cluster$ \textbf{in parallel}} 
            \FOR {$batch\_size = 1,2,4,8,\cdots, mbs_i$}
                \STATE model.step($batch\_size$)
                \STATE $p_i.append($\textit{TimeConsumedDuringStep})
                \IF {find $OOM$ error}
                    \STATE $mbs_i \gets batch\_size - 1$
                \ENDIF
            \ENDFOR

            \STATE $low \gets mbs_i/2$
            \STATE $high \gets mbs_i$
            \WHILE{$low \leq high$}
                \STATE $mid \gets \lfloor{(low + high) / 2} \rfloor$
                \STATE $batch\_size_i \gets mid$
                \STATE model.step($batch\_size_i$)
                \STATE $p_i.append($\textit{TimeConsumedDuringStep})
                \IF {find $OOM$ error}
                    \STATE $high \gets mid$
                \ELSE
                    \STATE $low \gets mid$
                \ENDIF
            \ENDWHILE
            \STATE $mbs_i \gets batch\_size_i$
        \ENDFOR
         
    \end{algorithmic}
\end{algorithm}

\subsection{Online Profiling of GPU}
In this section, Poplar evaluates the compute capability and memory capacity of each GPU simultaneously. Max batch size is an important factor for GPU utilization and amount of memory used in the GPU. Large micro-batch size can improve the GPU utilization and FLOPs, but it determining the maximum batch size is not straightforward. Researchers often need to manually adjust the batch size to meet the global batch size requirement and ensure no OOM error occurs before starting training. This is already a cumbersome task for homogeneous clusters, and it becomes even more complex for heterogeneous clusters as the varying memory capacities of GPUs. Therefore, Poplar achieves fully automated determination of the maximum batch size of each GPU. Besides, Poplar obtains performance data by collecting runtime metrics during online profiling.

\subsubsection{Heterogeneity Aware}
Accurately measuring the memory capacity of each GPU is challenging. The first challenge is how to ensure that the identified $mbs$ does not result in out-of-memory(OOM) errors during subsequent training phases. Previous work always use a cost model to estimate the memory capacity, and reserve a large amount of memory to prevent OOM errors, which would reduce GPU utilization. This is particularly unacceptable in a heterogeneous GPU cluster since some lower-end GPUs already own very limited memory. Another challenge is the search cost, which becomes extremely high when we try every possible number of batch.

Based on these, Poplar use two steps to search $mbs$ for each GPU, as shown in Algorithm \ref{power}. Poplar use two methods to accelerate the progress correspond to the two \textbf{for} loops in Algorithm 1. Firstly, given the linear relationship between the batch size and memory consumption during model training, Poplar initially estimate the $mbs$ under the condition of one batch. The $mbs$ obtained at this step represents a theoretical maximum number. The actual $mbs$ on the GPU is typically lower than this value, necessitating a more precise determination. Poplar measures the compute performance of each GPU alone with the $mbs$ during next step. Secondly, Poplar runs the model multiple times, gradually increasing the batch size exponentially until it reaches the $mbs$ or encounter an OOM error. Then, Poplar employs a binary search to iteratively run the model until it finds the accurate $mbs$ that does not cause an OOM error. At the end of this phase, Poplar will obtain the $mbs$ for each GPU and its running time under several batch sizes. The overall time is acceptable because Poplar run only a single iteration, rather than the entire $gbs$. Meanwhile, starting from ZeRO-0, if Poplar find that the current stage cannot even run a single batch, it will automatically increase the ZeRO stage. 

\subsubsection{Time Consumed Estimation}
In online profiling, the overall process across different ZeRO stages is almost the same, with the only difference lying in the computation of \textit{TimeConsumedDuringStep}, which records the GPU runtime during model execution. The most time-consuming operations are typically tensor multiplications and additions during the model computation process, but other operations cannot be neglected, such as tensor transpositions, activations, normalizations, convolution matrix inversions, eigenvalue decompositions, and memory accesses\cite{zhu2018benchmarking}. Therefore, we gather the wall time of each GPU to measuring compute capability instead of FLOPs alone. Besides, the communication occurs in Poplar among GPUs using Collective Operations\cite{thakur2005optimization} such as All-Reduce and All-Gather\cite{patarasuk2009bandwidth}. Therefore, we only need to consider the heterogeneity of compute performance due to the consistent communication time.

In ZeRO-0, the synchronization point occurs just before the optimizer updates the parameters, following the completion of both the forward and backward passes on all GPUs. At this point, the model needs to collect and accumulate gradients across all GPUs. The optimizer update time is very short, and even equal when using parameter offloading. Thus, Poplar record the total GPU runtime for both the forward and backward in ZeRO-0, which serves as the result for \textit{TimeConsumedDuringStep}. In ZeRO-1, the situation is identical to ZeRO-0 until the step of optimizer, where a second synchronization point is encountered. Therefore, Poplar use the same \textit{TimeConsumedDuringStep} method as in ZeRO-0.

In ZeRO-2, the first synchronization point occurs in backward. During the entire backward process, multiple synchronizations take place, with each synchronization using Reduce-Scatter to distribute the gradients to the GPUs responsible for maintaining the corresponding weights. The large number of synchronizations make the situation complex due to the overlap between computation and communication. Therefore, Poplar uses a noval method based on the communicate time. In each iteration, all GPUs are synchronized at the beginning and end of the backward, ensuring that the backward time is consistent across GPUs. However, the synchronization time is inconsistent, with faster GPUs doing Collective Operations earlier and consequently start to wait. The idle time is included in the time of Collective Operations. Thus, Poplar obtains the computation time during the backward process for each GPU by directly subtracting the time of Collective Operations, and obtain \textit{TimeConsumedDuringStep} by adding the time of forward. 

In ZeRO-3, there are multiple synchronization points during the whole pass. An extra synchronization is performed compared to ZeRO-2.  ZeRO-3 uses all-gather to collect weights before each computation. Thus, Poplar calculate the GPU runtime by subtracting three parts: (1)the time of All-Gather in the forward process, (2)the time of All-Gather and (3)Reduce-Scatter in the backward process.

\subsection{Offline Analyzing of Configuration}
At this stage, Poplar first construct comprehensive performance curves which contain the computation times for each GPU across all batch sizes. Subsequently, Poplar uses a search algorithm to determine the optimal batch allocation strategy. The factors in search algorithm include compute power, memory capacity, the number of each type of GPU and network communication time. Finally, Poplar completes preparatory tasks for model initialization, such as configuring a data loader with dynamic batch sizes, and distributing model shards to each GPU. Poplar keep the global batch size consistent and the micro batch size dynamic, which has a negligible impact of batch-related operations(e.g., batch normalization\cite{bjorck2018understanding}).

\begin{algorithm}[tb]
    \renewcommand{\algorithmicrequire}{\textbf{Input:}}
	\renewcommand{\algorithmicensure}{\textbf{Output:}}
	\caption{Optimal Batch Size Searching}
    \label{alloc}
    \begin{algorithmic}[1]
        \REQUIRE  global batch size: $gbs$,\\ compute capability: $\mathbf{P} = \{p_1, p_2, \cdots, p_n\}$
	    \ENSURE Configuration: $\mathbf{B} = \{b_1, b_2, \cdots, b_n\}$

        \IF {$ZeRO\_stage == 0$ or $1$}
            \FOR{$i = 1,2,\cdots,n$} 
                \STATE $speed_i \gets max(p_i)$
            \ENDFOR
            \STATE $speed_{cluster} \gets \sum_{i=1}^n speed_i$
            \STATE $time_{optimal} \gets gbs/speed_{cluster} $
            \FOR{$i = 1,2,\cdots,n$} 
                \STATE $gmbs \gets \lfloor{time_{optimal} \times speed_i} \rfloor$
                \STATE $\delta_i \gets time_{optimal} / speed_i$
                \STATE $u_i \gets \delta_i \times speed_i$
            \ENDFOR
            \STATE $batch_{remain} \gets gbs-sum(\mathbf{B})$
            \WHILE {$batch_{remain} > 0$}
                \STATE $i \gets min(u_i)$
                \STATE $update(u_i, gmbs, batch_{remain})$
            \ENDWHILE
        \ELSIF{$ZeRO\_stage == 2$ or $3$}
            \FOR{$t $ in $ range(time_{min}, time_{max})$} 
                \FOR{$i = 1,2,\cdots,n$}
                    \STATE $batch\_size_i \gets find(g_i,t)$
                \ENDFOR
                \STATE $micro
                    \_batch\_size \gets \sum_{i=1}^n batch\_size_i$
                \STATE $gas \gets gbs / micro\_batch\_size$
                \STATE $wall\_time \gets (t+time\_communication) \times gas$  
                \IF {$wall\_time < optimal\_time$}
                    \STATE $optimal\_time \gets wall\_time$;
                    \STATE $b_i \gets batch\_size_i$
                \ENDIF
            \ENDFOR
        \ENDIF
    \end{algorithmic}
\end{algorithm}

\subsubsection{Optimal Batch Size Searching}
The choice of the batch size significantly affects throughput of GPU\cite{narayanan2021efficient}. During the process of increasing the batch size from 0 to $mbs$, the GPU's speed for computing one batch data initially rises rapidly, then gradually levels off, eventually reaching a plateau where further increases in batch size do not improve computational speed. We conducted extensive experiments to validate this trend, and the results confirm that the relationship between batch size and compute capability of different GPU is universal. The detail of above experiments are provided in Appendix.

Based on this observation, Poplar constructs complete performance curves with $p_i$. Initially, Poplar divides each $TimeConsumedDuringStep$ in $p_i$ by the corresponding batch size to determine the speed in some discrete points. Then, Poplar employs cubic spline interpolation\cite{mckinley1998cubic} to fit the computing performance data for each GPU. Cubic spline interpolation approximates the original data by constructing a series of local cubic polynomials that join smoothly at each data point, ensuring both continuity and smoothness. In our work, the data obtained after Online Profiling exhibit characteristics well-suited to cubic spline interpolation, which also needs low computational complexity. We provide a detailed mathematical formulation of cubic spline interpolation in Appendix and analyze the error between the interpolated curves and the actual performance data. After obtaining the complete curve, Poplar can determine the batch size range that ensures each GPU operates at its peak performance. During the search process, Poplar aims to allocate batches within these ranges for each GPU. Given that Poplar treats each card independently, the performance of the cluster can be directly calculated through the sum of each GPU's data. The process is detailed in Algorithm \ref{alloc}.

In ZeRO-1 and ZeRO-2, the search algorithm needs to ensure that all nodes complete their computations simultaneously after backward. Consequently, Poplar directly allocates $gbs$ based on each GPU's peak computational capacity, and determines the global micro batch size ($gmbs$) for each GPU. The $gbs$ represents the amount of data each GPU needs to compute in a single itreation through gradient accumulation\cite{soydaner2020comparison} before the first synchronization. The sum of all GPUs' $gmbs$ equals $gbs$. Then, Poplar assigns a suitable $b_i$ to each GPU, ensuring that $b_i$ falls within the range that maximizes the GPU's compute capability. Since the allocation progress is an integer programming problem, there will be some $remain\_batch$ that remain unallocated. Therefore, in the second step, Poplar uses equation (\ref{delta time}) and (\ref{under utilization}) to iteratively allocate $remain\_batch$ to nodes with relatively lower workload, resulting in the last batch size ($lbs$) for gradient accumulation. $lbs$ is the final iteration for the gradient accumulation loop on each GPU.

In ZeRO-2 and ZeRO-3, the situation differs due to the additional communication. The search algorithm need to consider both load balancing and communication overhead. Selecting a smaller $micro\_batch\_size$ requires a larger number of gradient accumulation steps to complete $gbs$, leading to a increase in communication time. Conversely, a larger $micro\_batch\_size$ may increase the imbalance in workload, thus increasing idle times and decreasing cluster utilization. During the search process, Poplar explores every possible value of $t$, which represents the time it takes for the entire cluster to complete a data of $micro\_batch\_size$. Each value of $t$ corresponds to a potential configuration. Poplar calculates the time required to complete $gbs$ and identifies a configuration $B$ that minimizes the wall time by iterating try a large amount of $t$. The wall time can be calculated ad $t$ plus the communication time per step $time\_communication$, and then multiplied by the total number of gradient accumulation steps. The function $find$ is responsible for determining the maximum batch size GPU can compute given the current value of $t$. 

\begin{figure*}[t]
\centering
\includegraphics[width=2\columnwidth]{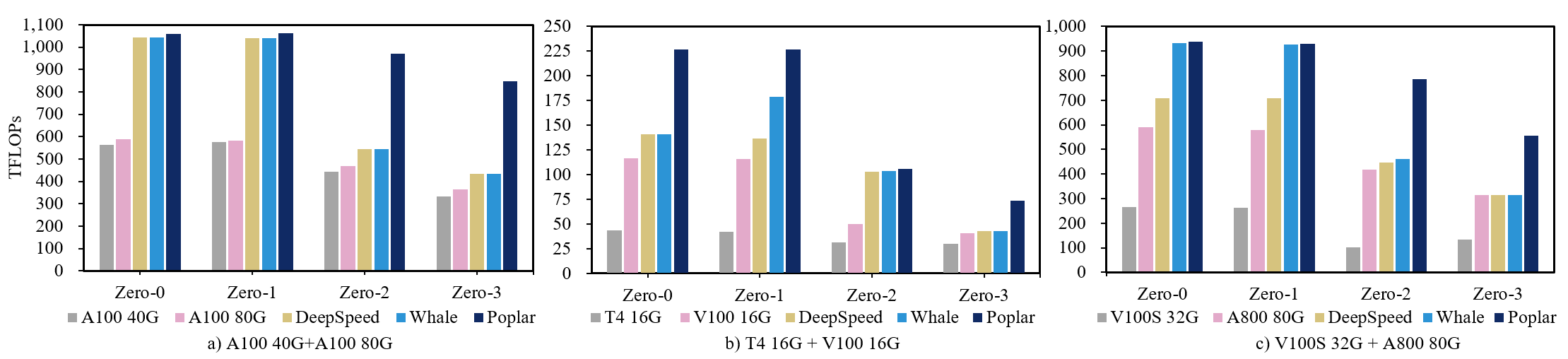}
\caption{The performance on three types of heterogeneous environments. Poplar performs better than all baselines.}
\label{Performance experiment}
\end{figure*}

The use of the last batch size and gradient accumulation steps, along with independent allocation for each GPU, allows Poplar to overcome the traditional constraints between global batch size and micro batch size, thereby supporting the heterogeneity of quantity. This ability enables the workload to be distributed across GPUs in a fine-grained manner, thereby minimizing GPU idle times. Poplar strives to select larger batch sizes for each GPU to reduce overall communication, as network links between devices in heterogeneous clusters often presents a bottleneck. Besides, we modify the dataloader to support the use of $lbs$ and $gabs$ in Poplar. 

\section{Experiments}

\subsection{Setup}
We have implemented our work on PyTorch with around 2000+ lines of code. Poplar is decoupled from model training, thus any techniques that enhance LLM training efficiency can be seamlessly integrated through just a few lines of code, such as pipeline parallelism\cite{huang2019gpipe, narayanan2019pipedream}, parameter offloading\cite{ren2021zero, rajbhandari2021zero}, ZeRO++\cite{wang2023zero++}. 

Our experiments are conducted on three heterogeneous GPU clusters, each cluster contains two types of GPUs, as shown in Table \ref{clusters}. All experiments are evaluated on wikitext-2-v1 dataset\cite{merity2016pointer}. In our experiments, we use TFLOPs (FLOPs/1e12) as the metric for evaluation end-to-end utilization of cluster. For each experiment, we conduct 50 iterations and use the average value.

\subsection{Models and Baselines}
In the main experiments, we use Llama\cite{touvron2023llama} to validate performance. Subsequently, we employ Llama and BERT\cite{devlin-etal-2019-bert} of varying sizes to evaluate generality. We maintain a global batch size of 2 million tokens throughout our experiments. To clearly illustrate Poplar's acceleration capabilities, we establish four baselines. Baseline 1 uses less powerful homogeneous GPUs, while baseline 2 uses more powerful homogeneous GPUs. The third baseline employs DeepSpeed, a state-of-the-art distributed training system. We  manually allocate maximum batch sizes that meet the constraints for baseline 3. The fourth baseline is Whale \cite{jia2022whale}, a state-of-the-art heterogeneous training system that supports hetero-aware load balancing. In baseline 4, we also manually configure the maximum batch sizes to be consistent with its strategy.

\begin{table}[ht]
    \centering
    \begin{tabular}{cccc}
        \toprule
        Cluster & GPU & Number & Intra-Link\\ \midrule
        A & \begin{tabular}[c]{@{}c@{}}A100 80GB\\ A100 40GB\end{tabular} & \begin{tabular}[c]{@{}c@{}}4\\ 4\end{tabular} & \begin{tabular}[c]{@{}c@{}}NVLink\\ PCIE\end{tabular}\\ \midrule
        B & \begin{tabular}[c]{@{}c@{}}V100 16GB\\ T4 16GB\end{tabular} & \begin{tabular}[c]{@{}c@{}}2\\ 2\end{tabular} & PCIE \\ \midrule
        C & \begin{tabular}[c]{@{}c@{}}A800 80GB\\ V100S 32GB\end{tabular} & \begin{tabular}[c]{@{}c@{}}4\\ 4\end{tabular} & PCIE \\ 
        \bottomrule
    \end{tabular}
\caption{Three clusters in our experiments, each cluster contains two types of GPUs. Number represents the quantity of corresponding GPUs in the cluster, while Inter-Link denotes the networking connection between GPUs.}
\label{clusters}
\end{table}

\begin{figure*}[t]
\centering
\includegraphics[width=2\columnwidth]{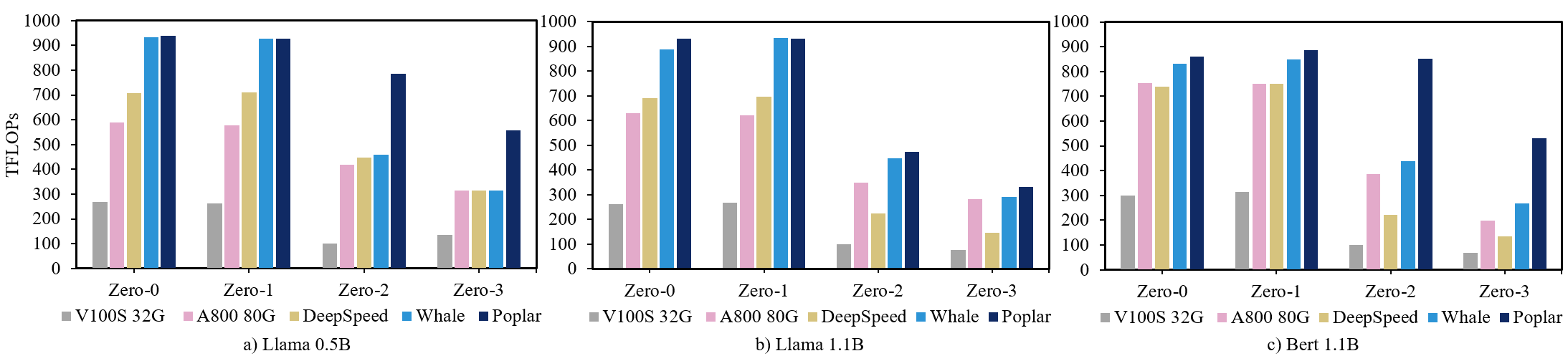}
\caption{Results on different models. Poplar performs better on BERT compared to Llama. Due to a small micro batch size, Poplar performs less well at 1.1B parameters than 0.5B parameters.}
\label{models experiment}
\end{figure*}

\begin{figure}[ht]
\centering
\includegraphics[width=1\columnwidth]{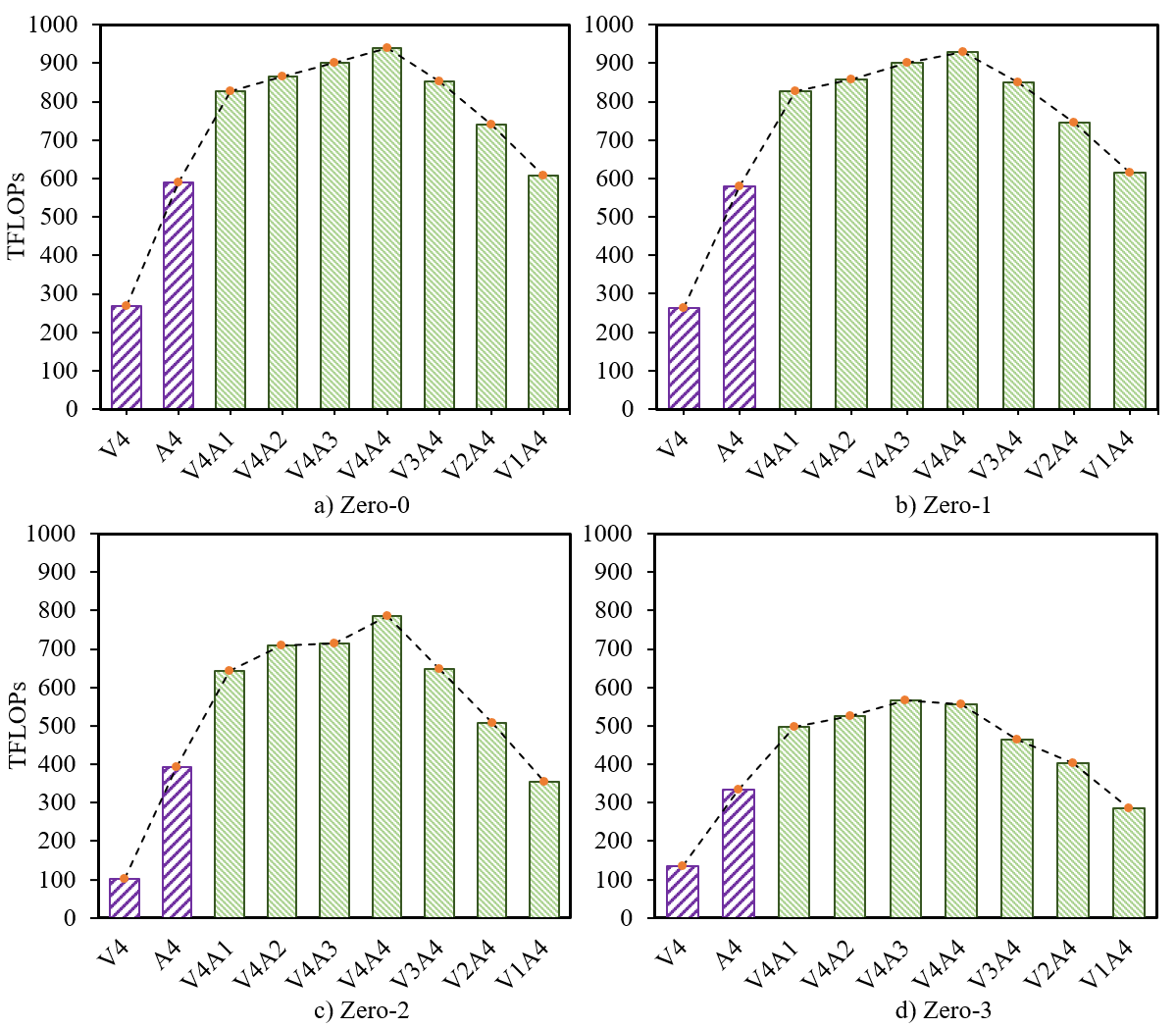}
\caption{The evaluation on Poplar's training capabilities across varying numbers of heterogeneous GPUs. The numbers and letters in the figure indicate the quantity of corresponding GPUs, for example, V4 denotes four V100S, A4 denotes four A800, and V4A1 denotes four V100S with one A800.}
\label{Quantity experiment}
\end{figure}

\subsection{Performance}
In the main experiments, we comprehensive verify and compare Poplar's performance across three types of heterogeneous environments. To better illustrate the differences among all heterogeneous GPUs and Poplar's acceleration in various clusters, we set the model's parameter to 0.5 billion, ensuring that all baselines can run. As shown in Figure \ref{Performance experiment}, the three clusters represent three distinct heterogeneous scenarios: (1)different memory capacities with equal compute capabilities, (2)different compute capabilities with equal memory capacities, and (3)different memory capacities with different compute capabilities. Figures \ref{Performance experiment}(a), \ref{Performance experiment}(b), and \ref{Performance experiment}(c) display the performance of clusters A, B, and C, respectively. Within each subplot, the five baselines are presented from left to right for each ZeRO stage. 

The experimental results demonstrate that Poplar performs well across all scenarios. The improvements in ZeRO-0 and ZeRO-1, as shown in Figures \ref{Performance experiment}(a) and \ref{Performance experiment}(c), arise from a larger $gmbs$ to A100 80G and A800 80G. Notably, despite having the same FLOPs rating for A100 40G and A100 80G, the A100 80G supports a larger $mbs$, leading to a improvement per iteration. In cluster A, Whale performs similarly to DeepSpeed, which lacks heterogeneity awareness due to the homogeneous matrix computation capabilities of the two kind of GPUs. The enhancements in ZeRO-0 and ZeRO-1 depicted in Figure \ref{Performance experiment}(b) mainly come from the finer-grained aware of GPU compute capability, indicating that, under certain heterogeneous hardware conditions, the overhead of components other than matrix multiplication should not be overlooked. The improvements in ZeRO-2 and ZeRO-3 are substantial across all three cases, as Poplar effectively reduces the number of gradient accumulation steps, thereby decreasing communication volume. This ability is crucial for training on heterogeneous clusters, where the network may encounter a bottleneck. The ablation study of Poplar is given in Appendix.

\subsubsection{Different Quantity}
In the second experiment, we evaluate Poplar's training capabilities across varying numbers of heterogeneous GPUs. We select cluster C to evaluate the support for arbitrary numbers of heterogeneous GPUs, with both differing memory capacities and compute capabilities. As shown in Figure \ref{Quantity experiment}, each subplot includes three sections of data: the first section uses only the weaker V100S, the second section uses only the stronger A800, and the third section uses different proportions of the two heterogeneous GPU types, including ratios of 4:1, 4:2, 4:3, 4:4, 3:4, 2:4, and 1:4 for A800 to V100S. From the figure, it is clear that as the number of GPUs increases, Poplar achieves progressively higher cluster performance. Additionally, it is evident that a reduction in the number of A800 leads to a more significant decrease in overall cluster performance compared to a reduction in the number of V100S.

\subsubsection{Different models}
Heterogeneous clusters are highly sensitive to model size and structure. Therefore, we conducted additional experiments to demonstrate Poplar's acceleration capabilities on heterogeneous clusters. We used a 1.1B Llama model and a 1.1B BERT model for comparison with the 0.5B Llama model, as shown in Figure \ref{models experiment}. In Figure \ref{models experiment}(b), Poplar achieves up to a 2.27x improvement over DeepSpeed and up to a 1.14x improvement over Whale. In Figure \ref{models experiment}(c), Poplar demonstrates even more substantial cluster performance improvements, achieving up to a 3.92x improvement over DeepSpeed and up to a 1.98x improvement over Whale.

\subsection{Time Overhead}
Poplar achieves performance gains through fine-grained planning before the start of training. In this section, we analyze the time overhead incurred during Poplar's preliminary analysis phase. The majority of the time overhead is concentrated in Online Profiling, as Poplar requires multiple runs of the model to aware more accurate heterogenity. However, this approach eliminates the need for researchers to manually find suitable batch sizes. The runtime of the Online Profiling phase can vary depending on several factors, including the state of the GPUs, network links, ZeRO stage, model size and architecture. The Offline Analysis phase primarily handles numerical calculations and incurs relatively minimal time overhead. We detail the time overhead for our main experiment in Appendix.

\section{Conclusion}
In this paper, we propose a distributed training system which extends each ZeRO stage with heterogeneous-aware capabilities. Our system can adapt to a broader range of heterogeneous environments by dynamically assign batches to each GPU independently. In the experiments, we show that our system outperforms state-of-the-art training systems.
In the future, we will explore the following directions:
\begin{enumerate}
    \item We will consider unevenly distributing model parameters across heterogeneous devices based on their memory sizes in different ZeRO stages.
    \item We will further study the application of ZeRO in network-constrained heterogeneous clusters.
\end{enumerate}
\bibliography{aaai25}

\begin{thebibliography}{36}
\providecommand{\natexlab}[1]{#1}

\bibitem[{Bjorck et~al.(2018)Bjorck, Gomes, Selman, and Weinberger}]{bjorck2018understanding}
Bjorck, N.; Gomes, C.~P.; Selman, B.; and Weinberger, K.~Q. 2018.
\newblock Understanding batch normalization.
\newblock \emph{Advances in neural information processing systems}, 31.

\bibitem[{Cao, Zhu, and Zhou(2021)}]{cao2021sap}
Cao, J.; Zhu, Z.; and Zhou, X. 2021.
\newblock SAP-SGD: Accelerating distributed parallel training with high communication efficiency on heterogeneous clusters.
\newblock In \emph{2021 IEEE International Conference on Cluster Computing (CLUSTER)}, 94--102. IEEE.

\bibitem[{Chen et~al.(2021)Chen, Weng, Wang, Li, and Li}]{chen2021accelerating}
Chen, C.; Weng, Q.; Wang, W.; Li, B.; and Li, B. 2021.
\newblock Accelerating distributed learning in non-dedicated environments.
\newblock \emph{IEEE Transactions on Cloud Computing}, 11(1): 515--531.

\bibitem[{Choquette(2022)}]{choquette2022nvidia}
Choquette, J. 2022.
\newblock Nvidia hopper gpu: Scaling performance.
\newblock In \emph{2022 IEEE Hot Chips 34 Symposium (HCS)}, 1--46. IEEE Computer Society.

\bibitem[{Devlin et~al.(2019)Devlin, Chang, Lee, and Toutanova}]{devlin-etal-2019-bert}
Devlin, J.; Chang, M.-W.; Lee, K.; and Toutanova, K. 2019.
\newblock {BERT}: Pre-training of Deep Bidirectional Transformers for Language Understanding.
\newblock In Burstein, J.; Doran, C.; and Solorio, T., eds., \emph{Proceedings of the 2019 Conference of the North {A}merican Chapter of the Association for Computational Linguistics: Human Language Technologies, Volume 1 (Long and Short Papers)}, 4171--4186. Minneapolis, Minnesota: Association for Computational Linguistics.

\bibitem[{Geng et~al.(2023)Geng, Cao, Jia, Zhu, Fang, Gao, Ji, Jia, Han, and Zhou}]{geng2023heter}
Geng, J.; Cao, J.; Jia, H.; Zhu, Z.; Fang, H.; Gao, C.; Ji, C.; Jia, G.; Han, G.; and Zhou, X. 2023.
\newblock Heter-Train: A Distributed Training Framework Based on Semi-Asynchronous Parallel Mechanism for Heterogeneous Intelligent Transportation Systems.
\newblock \emph{IEEE Transactions on Intelligent Transportation Systems}, 25(1): 959--972.

\bibitem[{Gerbessiotis and Valiant(1994)}]{gerbessiotis1994direct}
Gerbessiotis, A.~V.; and Valiant, L.~G. 1994.
\newblock Direct bulk-synchronous parallel algorithms.
\newblock \emph{Journal of parallel and distributed computing}, 22(2): 251--267.

\bibitem[{Huang et~al.(2019)Huang, Cheng, Bapna, Firat, Chen, Chen, Lee, Ngiam, Le, Wu et~al.}]{huang2019gpipe}
Huang, Y.; Cheng, Y.; Bapna, A.; Firat, O.; Chen, D.; Chen, M.; Lee, H.; Ngiam, J.; Le, Q.~V.; Wu, Y.; et~al. 2019.
\newblock Gpipe: Efficient training of giant neural networks using pipeline parallelism.
\newblock \emph{Advances in neural information processing systems}, 32.

\bibitem[{Jia et~al.(2022)Jia, Jiang, Wang, Xiao, Shi, Zhang, Li, Chen, Li, Zheng et~al.}]{jia2022whale}
Jia, X.; Jiang, L.; Wang, A.; Xiao, W.; Shi, Z.; Zhang, J.; Li, X.; Chen, L.; Li, Y.; Zheng, Z.; et~al. 2022.
\newblock Whale: Efficient giant model training over heterogeneous $\{$GPUs$\}$.
\newblock In \emph{2022 USENIX Annual Technical Conference (USENIX ATC 22)}, 673--688.

\bibitem[{Jia, Zaharia, and Aiken(2019)}]{jia2019beyond}
Jia, Z.; Zaharia, M.; and Aiken, A. 2019.
\newblock Beyond data and model parallelism for deep neural networks.
\newblock \emph{Proceedings of Machine Learning and Systems}, 1: 1--13.

\bibitem[{Kwon et~al.(2020)Kwon, Yu, Jeong, and Chun}]{kwon2020nimble}
Kwon, W.; Yu, G.-I.; Jeong, E.; and Chun, B.-G. 2020.
\newblock Nimble: Lightweight and parallel gpu task scheduling for deep learning.
\newblock \emph{Advances in Neural Information Processing Systems}, 33: 8343--8354.

\bibitem[{Lai et~al.(2023)Lai, Li, Tang, Ge, Liu, Duan, Qiao, and Li}]{lai2023merak}
Lai, Z.; Li, S.; Tang, X.; Ge, K.; Liu, W.; Duan, Y.; Qiao, L.; and Li, D. 2023.
\newblock Merak: An efficient distributed dnn training framework with automated 3d parallelism for giant foundation models.
\newblock \emph{IEEE Transactions on Parallel and Distributed Systems}, 34(5): 1466--1478.

\bibitem[{Li et~al.(2022)Li, Wang, Xing, and Zhang}]{li2022amp}
Li, D.; Wang, H.; Xing, E.; and Zhang, H. 2022.
\newblock Amp: Automatically finding model parallel strategies with heterogeneity awareness.
\newblock \emph{Advances in Neural Information Processing Systems}, 35: 6630--6639.

\bibitem[{Luo et~al.(2019)Luo, Lin, Zhuo, and Qian}]{luo2019hop}
Luo, Q.; Lin, J.; Zhuo, Y.; and Qian, X. 2019.
\newblock Hop: Heterogeneity-aware decentralized training.
\newblock In \emph{Proceedings of the Twenty-Fourth International Conference on Architectural Support for Programming Languages and Operating Systems}, 893--907.

\bibitem[{McKinley and Levine(1998)}]{mckinley1998cubic}
McKinley, S.; and Levine, M. 1998.
\newblock Cubic spline interpolation.
\newblock \emph{College of the Redwoods}, 45(1): 1049--1060.

\bibitem[{Merity et~al.(2016)Merity, Xiong, Bradbury, and Socher}]{merity2016pointer}
Merity, S.; Xiong, C.; Bradbury, J.; and Socher, R. 2016.
\newblock Pointer Sentinel Mixture Models.
\newblock arXiv:1609.07843.

\bibitem[{Narayanan et~al.(2019)Narayanan, Harlap, Phanishayee, Seshadri, Devanur, Ganger, Gibbons, and Zaharia}]{narayanan2019pipedream}
Narayanan, D.; Harlap, A.; Phanishayee, A.; Seshadri, V.; Devanur, N.~R.; Ganger, G.~R.; Gibbons, P.~B.; and Zaharia, M. 2019.
\newblock PipeDream: Generalized pipeline parallelism for DNN training.
\newblock In \emph{Proceedings of the 27th ACM symposium on operating systems principles}, 1--15.

\bibitem[{Narayanan et~al.(2021)Narayanan, Shoeybi, Casper, LeGresley, Patwary, Korthikanti, Vainbrand, Kashinkunti, Bernauer, Catanzaro et~al.}]{narayanan2021efficient}
Narayanan, D.; Shoeybi, M.; Casper, J.; LeGresley, P.; Patwary, M.; Korthikanti, V.; Vainbrand, D.; Kashinkunti, P.; Bernauer, J.; Catanzaro, B.; et~al. 2021.
\newblock Efficient large-scale language model training on gpu clusters using megatron-lm.
\newblock In \emph{Proceedings of the International Conference for High Performance Computing, Networking, Storage and Analysis}, 1--15.

\bibitem[{Park et~al.(2020)Park, Yun, Chang, Nguyen, Lee, Choi, Noh, and Choi}]{park2020hetpipe}
Park, J.~H.; Yun, G.; Chang, M.~Y.; Nguyen, N.~T.; Lee, S.; Choi, J.; Noh, S.~H.; and Choi, Y.-r. 2020.
\newblock $\{$HetPipe$\}$: Enabling large $\{$DNN$\}$ training on (whimpy) heterogeneous $\{$GPU$\}$ clusters through integration of pipelined model parallelism and data parallelism.
\newblock In \emph{2020 USENIX Annual Technical Conference (USENIX ATC 20)}, 307--321.

\bibitem[{Patarasuk and Yuan(2009)}]{patarasuk2009bandwidth}
Patarasuk, P.; and Yuan, X. 2009.
\newblock Bandwidth optimal all-reduce algorithms for clusters of workstations.
\newblock \emph{Journal of Parallel and Distributed Computing}, 69(2): 117--124.

\bibitem[{Rajbhandari et~al.(2020)Rajbhandari, Rasley, Ruwase, and He}]{rajbhandari2020zero}
Rajbhandari, S.; Rasley, J.; Ruwase, O.; and He, Y. 2020.
\newblock Zero: Memory optimizations toward training trillion parameter models.
\newblock In \emph{SC20: International Conference for High Performance Computing, Networking, Storage and Analysis}, 1--16. IEEE.

\bibitem[{Rajbhandari et~al.(2021)Rajbhandari, Ruwase, Rasley, Smith, and He}]{rajbhandari2021zero}
Rajbhandari, S.; Ruwase, O.; Rasley, J.; Smith, S.; and He, Y. 2021.
\newblock Zero-infinity: Breaking the gpu memory wall for extreme scale deep learning.
\newblock In \emph{Proceedings of the international conference for high performance computing, networking, storage and analysis}, 1--14.

\bibitem[{Rasley et~al.(2020)Rasley, Rajbhandari, Ruwase, and He}]{rasley2020deepspeed}
Rasley, J.; Rajbhandari, S.; Ruwase, O.; and He, Y. 2020.
\newblock Deepspeed: System optimizations enable training deep learning models with over 100 billion parameters.
\newblock In \emph{Proceedings of the 26th ACM SIGKDD International Conference on Knowledge Discovery \& Data Mining}, 3505--3506.

\bibitem[{Ren et~al.(2021)Ren, Rajbhandari, Aminabadi, Ruwase, Yang, Zhang, Li, and He}]{ren2021zero}
Ren, J.; Rajbhandari, S.; Aminabadi, R.~Y.; Ruwase, O.; Yang, S.; Zhang, M.; Li, D.; and He, Y. 2021.
\newblock $\{$Zero-offload$\}$: Democratizing $\{$billion-scale$\}$ model training.
\newblock In \emph{2021 USENIX Annual Technical Conference (USENIX ATC 21)}, 551--564.

\bibitem[{Shoeybi et~al.(2019)Shoeybi, Patwary, Puri, LeGresley, Casper, and Catanzaro}]{shoeybi2019megatron}
Shoeybi, M.; Patwary, M.; Puri, R.; LeGresley, P.; Casper, J.; and Catanzaro, B. 2019.
\newblock Megatron-lm: Training multi-billion parameter language models using model parallelism.
\newblock arXiv:1909.08053.

\bibitem[{Smith et~al.(2022)Smith, Patwary, Norick, LeGresley, Rajbhandari, Casper, Liu, Prabhumoye, Zerveas, Korthikanti et~al.}]{smith2022using}
Smith, S.; Patwary, M.; Norick, B.; LeGresley, P.; Rajbhandari, S.; Casper, J.; Liu, Z.; Prabhumoye, S.; Zerveas, G.; Korthikanti, V.; et~al. 2022.
\newblock Using deepspeed and megatron to train megatron-turing nlg 530b, a large-scale generative language model.
\newblock arXiv:2201.11990.

\bibitem[{Song et~al.(2023)Song, Mi, Xie, and Chen}]{song2023powerinfer}
Song, Y.; Mi, Z.; Xie, H.; and Chen, H. 2023.
\newblock Powerinfer: Fast large language model serving with a consumer-grade gpu.
\newblock \emph{arXiv preprint arXiv:2312.12456}.

\bibitem[{Soydaner(2020)}]{soydaner2020comparison}
Soydaner, D. 2020.
\newblock A comparison of optimization algorithms for deep learning.
\newblock \emph{International Journal of Pattern Recognition and Artificial Intelligence}, 34(13): 2052013.

\bibitem[{Tang et~al.(2021)Tang, Du, Zhang, Yang, and Li}]{tang2021aeml}
Tang, Z.; Du, L.; Zhang, X.; Yang, L.; and Li, K. 2021.
\newblock AEML: An acceleration engine for multi-GPU load-balancing in distributed heterogeneous environment.
\newblock \emph{IEEE Transactions on Computers}, 71: 1344--1357.

\bibitem[{Thakur, Rabenseifner, and Gropp(2005)}]{thakur2005optimization}
Thakur, R.; Rabenseifner, R.; and Gropp, W. 2005.
\newblock Optimization of collective communication operations in MPICH.
\newblock \emph{The International Journal of High Performance Computing Applications}, 19(1): 49--66.

\bibitem[{Touvron et~al.(2023)Touvron, Martin, Stone, Albert, Almahairi, Babaei, Bashlykov, Batra, Bhargava, Bhosale et~al.}]{touvron2023llama}
Touvron, H.; Martin, L.; Stone, K.; Albert, P.; Almahairi, A.; Babaei, Y.; Bashlykov, N.; Batra, S.; Bhargava, P.; Bhosale, S.; et~al. 2023.
\newblock Llama 2: Open foundation and fine-tuned chat models.
\newblock arXiv:2307.09288.

\bibitem[{Wang et~al.(2023)Wang, Qin, Jacobs, Holmes, Rajbhandari, Ruwase, Yan, Yang, and He}]{wang2023zero++}
Wang, G.; Qin, H.; Jacobs, S.~A.; Holmes, C.; Rajbhandari, S.; Ruwase, O.; Yan, F.; Yang, L.; and He, Y. 2023.
\newblock Zero++: Extremely efficient collective communication for giant model training.
\newblock arXiv:2306.10209.

\bibitem[{Weng et~al.(2022)Weng, Xiao, Yu, Wang, Wang, He, Li, Zhang, Lin, and Ding}]{weng2022mlaas}
Weng, Q.; Xiao, W.; Yu, Y.; Wang, W.; Wang, C.; He, J.; Li, Y.; Zhang, L.; Lin, W.; and Ding, Y. 2022.
\newblock $\{$MLaaS$\}$ in the wild: Workload analysis and scheduling in $\{$Large-Scale$\}$ heterogeneous $\{$GPU$\}$ clusters.
\newblock In \emph{19th USENIX Symposium on Networked Systems Design and Implementation (NSDI 22)}, 945--960.

\bibitem[{Zheng et~al.(2022)Zheng, Li, Zhang, Zhuang, Chen, Huang, Wang, Xu, Zhuo, Xing et~al.}]{zheng2022alpa}
Zheng, L.; Li, Z.; Zhang, H.; Zhuang, Y.; Chen, Z.; Huang, Y.; Wang, Y.; Xu, Y.; Zhuo, D.; Xing, E.~P.; et~al. 2022.
\newblock Alpa: Automating inter-and $\{$Intra-Operator$\}$ parallelism for distributed deep learning.
\newblock In \emph{16th USENIX Symposium on Operating Systems Design and Implementation (OSDI 22)}, 559--578.

\bibitem[{Zhou et~al.(2020)Zhou, Guo, Qu, Li, Li, Guo, and Wang}]{zhou2020petrel}
Zhou, Q.; Guo, S.; Qu, Z.; Li, P.; Li, L.; Guo, M.; and Wang, K. 2020.
\newblock Petrel: Heterogeneity-aware distributed deep learning via hybrid synchronization.
\newblock \emph{IEEE Transactions on Parallel and Distributed Systems}, 32(5): 1030--1043.

\bibitem[{Zhu et~al.(2018)Zhu, Akrout, Zheng, Pelegris, Jayarajan, Phanishayee, Schroeder, and Pekhimenko}]{zhu2018benchmarking}
Zhu, H.; Akrout, M.; Zheng, B.; Pelegris, A.; Jayarajan, A.; Phanishayee, A.; Schroeder, B.; and Pekhimenko, G. 2018.
\newblock Benchmarking and analyzing deep neural network training.
\newblock In \emph{2018 IEEE International Symposium on Workload Characterization (IISWC)}, 88--100. IEEE.

\end{thebibliography}

\clearpage

\section{Appendix}
\subsection{Analysis of Experiments}
Here are some other details analysis about our experiment. Firstly, it can be observed that Poplar does not show a substantial speed improvement over DeepSpeed and Whale in some of our experimental results. For instance, in the main experiments on $Cluster A$ using ZeRO-0 and ZeRO-1, the performance gain is small due to the relatively low computational load and short computation times. Besides, we iteratively tested and manually optimized the maximum batch size for DeepSpeed and Whale, which enhanced hardware utilization. This manual optimization process is laborious and would be impractical in larger heterogeneous clusters. Consequently, Poplar addresses this limitation by implementing an automated search mechanism.

We also observe that while using heterogeneous machines generally increases utilization compared to homogeneous clusters, the overall utilization might not increase linearly and could even decrease. For example, in the main experiment on $Cluster A$ using ZeRO-0, it is clear that Poplar's final cluster TFLOPs are lower than the sum of two homogeneous clusters. This can be attributed to two main reasons: (1)First, within a single physical node, GPU-to-GPU communication occurs via high-speed internal links such as NVLink and PCIe. However, between nodes, communication takes place through slower external links like InfiniBand (IB) or Sockets. When training with a combination of heterogeneous nodes, the slowest network connection becomes the bottleneck for the entire heterogeneous cluster, leading to decreased utilization. (2)Second, Poplar is built on top of the ZeRO, and as the number of devices increases, the overall communication volume grows. For instance, in ZeRO-3, the communication volume for a Feed-Forward Network (FFN) can be calculated as:
$$ All\_Gather = [(h \times 4h) + (4h \times h)] \times d $$
$$ Reduce\_Scatter = [(h \times 4h) + (4h \times h)] \times d $$
$$ Comm_{Forward} = All\_Gather$$
$$ Comm_{Backward} = All\_Gather + Reduce\_Scatter$$
$$ Comm_{Volume} = Comm_{Forward} + Comm_{Backward}$$
$$ Comm_{Volume} = 24dh^2$$
where h is the model's hidden size and 4h is the intermediate size. The communication volume in ZeRO-3 primarily occurs during the All-Gather operations in forward propagation and both the All-Gather and Reduce-Scatter operations in backward propagation. From this formula, it is evident that the communication volume is proportional to the number of devices. Therefore, simply increasing the number of heterogeneous devices does not guarantee an improvement in overall cluster performance, other factors must also be considered. Our experiments confirm this, as seen in the second experiment, where the V4A4 group has lower cluster utilization than the V4A3 group in ZeRO-3. Despite the increase in hardware components, performance declines as the added communication time outweighs the reduction in computation time with the introduction of an additional A800. This results in an overall decrease in performance, which is an area we plan to investigate further in our future work.

In our experiments, the global batch size ($gbs$) is not sufficiently large. For example, the $gbs$ is typically 4 million tokens in Llama, while we use 2 million tokens to enable comparisons across different clusters. Using a large $gbs$ in Cluster B would result in excessively long computation times. Poplar performs better with larger $gbs$, due to its precise heterogeneity-aware capabilities. 

\subsection{Details about ZeRO}
In the main body of our paper, we detailed several advantages that ZeRO offers over 3D Parallelism. Notably, our implementation system eliminates the need for expert-level knowledge, as it automatically determines the mini-batch size and manages hardware resources without human intervention. However, ZeRO has inherent limitations, particularly related to communication. 3D Parallelism supports more sophisticated designs, such as pipeline parallelism between nodes with low communication bandwidth, which can reduce overall communication overhead. In contrast, optimizing ZeRO requires additional techniques, such as Hierarchical Partitioning, Quantized Weight Communication, and Quantized Gradient Communication. We will explore these strategies in future work. In the rest of section, we enhance the analysis of different stages of ZeRO during the online profiling phase, based on descriptions provided in the original ZeRO and ZeRO++ papers.

ZeRO stage 0 (ZeRO-0) holds the full duplication of model states on each GPU, which is akin to the conventional DDP (Distributed Data Parallel). During ZeRO-0 training, gradients across all data parallel processes are averaged at the end of the backward propagation before computing the updates for the next step. The averaging is performed using an all-reduce communication collective. State-of-art implementation of all-reduce uses a two-step approach, where the first step is a reduce-scatter operation, which reduces different part of the data on different process. The next step is an all-gather operation where each process gathers the reduced data on all the process. The result of these two steps is an all-reduce. 

In ZeRO stage 1 (ZeRO-1), only optimizer states are split and spread across all GPUs in use. The processes of ZeRO-1 and ZeRO-0 are identical up until the point where the optimizer updates the parameters. Optimizers distributed across the individual GPU memories are responsible for updating their respective parameters. Following this, an all-gather operation is employed to ensure that all GPUs acquire the updated parameters from the current step, which are then utilized for the subsequent forward pass.

ZeRO stage 2 (ZeRO-2) partitions both optimizer states and gradients. With gradient partitioning, each process only stores the portion of the gradients, that is required to update its corresponding parameter partition. As such, instead of an all-reduce, ZeRO only requires a scatter-reduce operation on the gradients. After each process updates the partition of the parameters that it is responsible for, an all-gather is also performed to collect all the updated parameters in the same way as zero-1 does.

During model training, ZeRO-3 lazy-schedules the fetching of parameters until the computation needs to happen on a particular layer. Before forward propagation, ZeRO launches an all-gather to collect the full model weights and then computes the forward pass. Then ZeRO empties the all-gather weights buffer after forward computation completes. During backward, ZeRO re-collects all model weights again via a second all-gather to calculate gradients. Once gradients are calculated on each GPU, ZeRO empties weights buffer again and conducts a reduce-scatter operation to do gradient averaging and re-distribution. Model states and parameters are updated in optimizer step. 

ZeRO stage 3 (ZeRO-3) splits all three components of model states as parameters, gradients, and optimizer states. ZeRO-3 is the most memory efficient solution for model training at large scale, but at the cost of more collective communications. After parameter partitioning, each data parallel process only stores the parameters that it updates. Therefore, during the forward propagation it needs to receives the parameters for all the other partitions. However, this can be pipelined to avoid the memory overhead. Before computing the forward propagation on the part of the model corresponding to a particular partition, the data parallel process responsible for that partition can broadcast the weights to all the data parallel processes. Once the forward propagation for that partition is done, the parameters can be discarded. In other words, zero-3 reschedule the parameter all-gather by spreading it across the entire forward propagation, and discarding the parameters once they have been used. Note however that this all-gather needs to happen once again for the backward propagation in the reverse order. Once gradients are calculated on each GPU, ZeRO-3 conducts a reduce-scatter operation to do gradient averaging and re-distribution. Finally, the optimizer only needs to update the portion of the parameters that it maintains.

\subsection{Relationship Between Compute Capability and Batch Size}
\begin{figure}[ht]
\centering
\includegraphics[width=1\columnwidth]{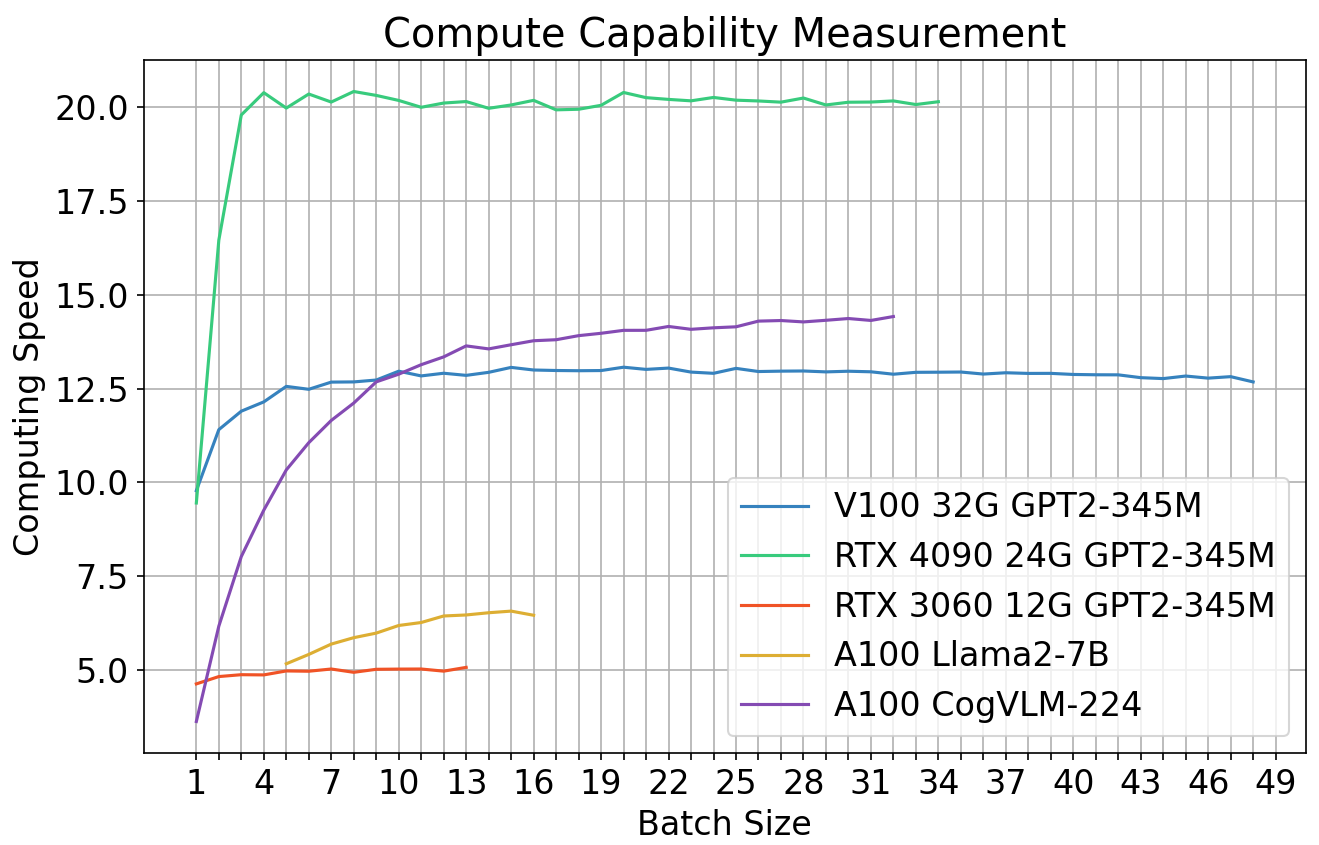} 
\caption{The results of the relationship between GPU compute capability and batch size across various conditions.}
\label{Realtionship}
\end{figure}

In this subsection, we enhance the analysis by testing the impact of batch size on GPU utilization. The results, shown in Figure \ref{Realtionship}, align with the evaluation data provided by NVIDIA in their discussion of 'Typical Tile Dimensions in cuBLAS and Performance.' Although various tiling strategies exist, larger tiles offer more data reuse, allowing for reduced bandwidth usage and greater efficiency compared to smaller tiles. However, for a given problem size, using larger tiles results in fewer tiles running in parallel, which can potentially lead to under-utilization of the GPU.

In our experiments, we use five distinct configurations, including four different GPUs, two of which are consumer-grade: (1) the GeForce RTX 4090 24G, (2) the GeForce RTX 3060 12G, (3) the V100 32G, and (4) the A100 80G. We test three models: (1) GPT-2 with 345 million parameters, (2) Llama with 7 billion parameters, and (3) the multimodal model CogVLM-224. Our experiments employ various training frameworks: Megatron-LM, BMTrain, and Hugging Face Accelerate. The results confirm the curve relationships discussed in the main text. Note that we record the time for a single run without averaging over multiple runs, which results in noticeable fluctuations in some cases.

\begin{figure*}[!th]
\centering
\includegraphics[width=2\columnwidth]{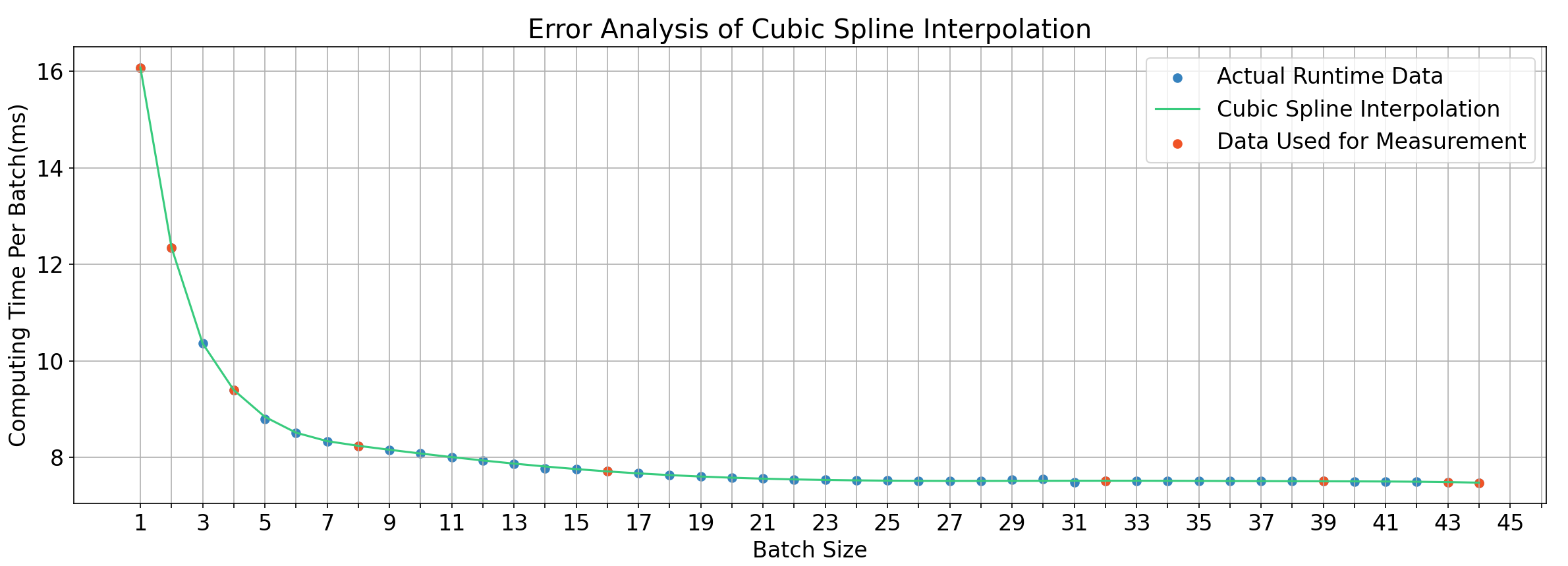} 
\caption{The gap between the data generated by Cubic Spline Interpolation and the actual operational data is almost zero.}
\label{CSI}
\end{figure*}

\subsection{Analysis of Compute Capability Measurement Methods}
This subsection highlights the limitations of using FLOPs as the sole metric for load balancing in heterogeneous GPU clusters. In our experiment, we run the 0.5B Llama model, excluding communication operations. Each GPU performs five iterations at its respective $mbs$, and we compute the average performance. To assess the relative performance of different GPUs, we measure the compute capability of each GPU and normalize it by dividing it by the compute capability of T4. This approach yields a relative compute capability metric for direct comparison across all GPUs. Since matrix multiplication constitutes the largest portion of model execution time, we focus on testing each GPU’s $mbs$. When $mbs$ are not utilized simultaneously, the performance differences among heterogeneous GPUs become more pronounced, as illustrated by the peak range for high-end GPUs shown in Figure \ref{Realtionship}.

\begin{figure}[ht]
\centering
\includegraphics[width=1\columnwidth]{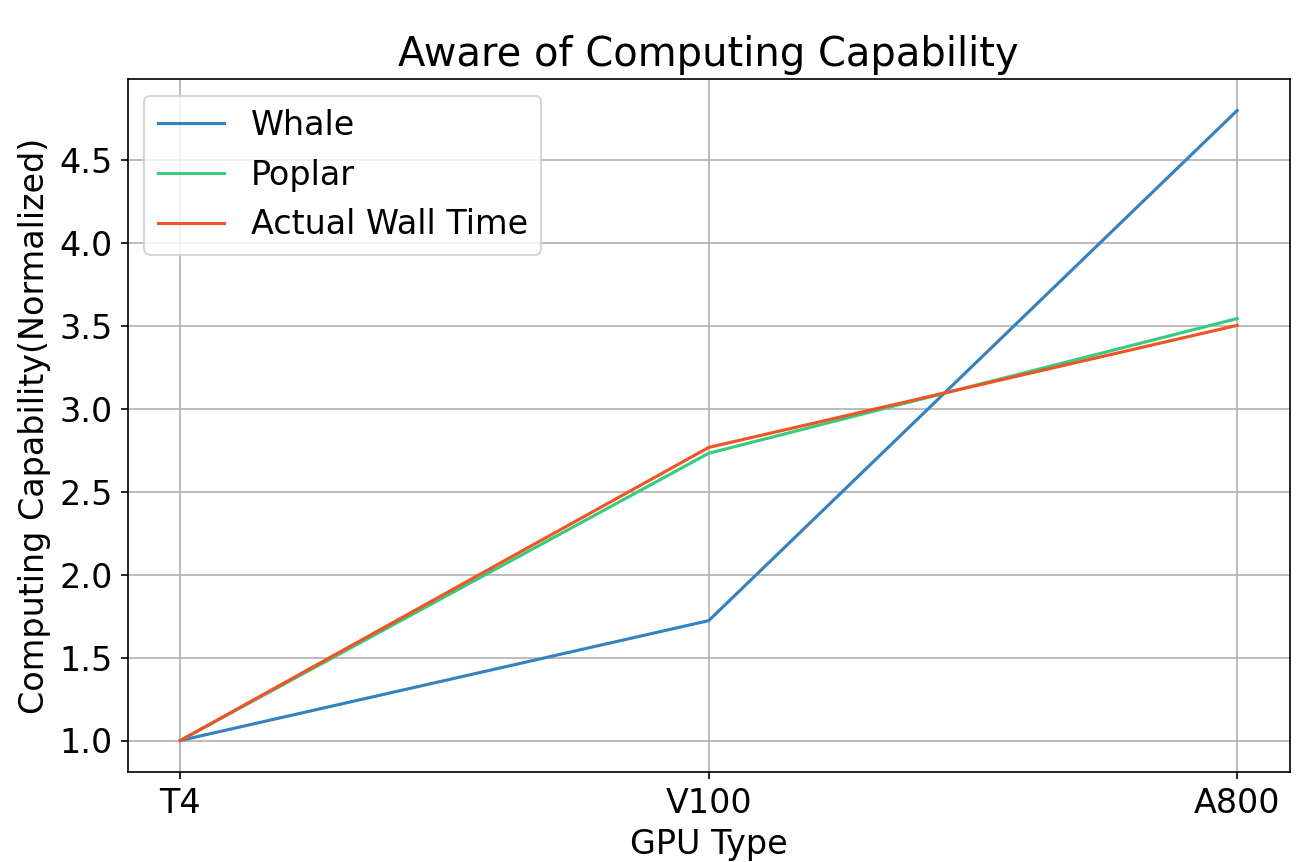} 
\caption{Compared to Whale, Poplar's compute 
capability measurement approach is significantly closer to the actual runtime conditions.}
\label{Aware Compute 
Capability}
\end{figure}

\subsection{Cubic Spline Interpolation}
Cubic spline interpolation is a method for constructing a smooth curve that passes through a given set of data points. This curve is composed of piecewise cubic polynomials, ensuring that the first and second derivatives are continuous across the data points. Cubic splines are often preferred over other forms of interpolation because they provide a good balance between smoothness and computational simplicity, avoiding the oscillations that can occur with higher-degree polynomial interpolations.

Given a set of data points \((x_i, y_i)\) for \(i = 0, 1, \ldots, n\), where the \(x_i\) are distinct, a cubic spline \(S(x)\) is a piecewise cubic polynomial that satisfies the following conditions.
\begin{enumerate}
    \item Piecewise Cubic Polynomials\\
    For each interval \([x_i, x_{i+1}]\) for \(i = 0, 1, \ldots, n-1\), \(S(x)\) is a cubic polynomial of the form:
    \[S_i(x) = a_i + b_i(x - x_i) + c_i(x - x_i)^2 + d_i(x - x_i)^3\]
    where \(a_i, b_i, c_i,\) and \(d_i\) are constants to be determined.
    \item Interpolation\\
    The spline passes through each data point, i.e., \(S(x_i) = y_i\) for \(i = 0, 1, \ldots, n\).
    \item Smoothness\\
    The first and second derivatives of \(S(x)\) are continuous at each interior data point \(x_i\), for \(i = 1, \ldots, n-1\):
    \[S'_i(x_{i+1}) = S'_{i+1}(x_{i+1}), \quad S''_i(x_{i+1}) = S''_{i+1}(x_{i+1})\]
    \item Boundary Conditions\\
    There are different types of boundary conditions that can be applied, such as natural cubic splines, which have zero second derivatives at the endpoints:
    \[S''_0(x_0) = S''_{n-1}(x_n) = 0\]
\end{enumerate}

The Cubic Spline Interpolation method is computationally efficient and exhibits very low error. As shown in Figure \ref{CSI}, we executed the 500M Llama model on the A800 80G and listed the discrete data obtained from actual runs across all batch sizes. We specifically highlighted a subset of these data points, which were directly utilized in Poplar for evaluating GPU computational performance. It can be observed that the generated curve closely matches the actual discrete data.

\begin{table}[ht]
    \centering
    \begin{tabular}{cccc}
        \toprule
        Stage & T4 & V100 & A800 \\ \midrule
        ZeRO-0 & 67s & 11s & 106s \\ \midrule
        ZeRO-1 & 137s & 19s & 143s \\ \midrule
        ZeRO-2 & 138s & 27s & 70s \\ \midrule
        ZeRO-3 & 174s & 31s & 71s \\ 
        \bottomrule
    \end{tabular}
\caption{Three clusters in our experiments, each cluster contains two types of GPUs. Number represents the quantity of corresponding GPUs in the cluster, while Inter-Link denotes the networking connection between GPUs.}
\label{Overhead}
\end{table}

\subsection{Overhead of Poplar}
In this section, we present time overhead data for three different GPUs across various ZeRO stages, covering all phases of Poplar. Several factors affect the time overhead, including GPU compute capability, memory capacity, PCIe speed, DRAM read/write speeds, CPU performance, and network performance. The binary search for $mbs$ often results in varying search paths, which can lead to inconsistent search times. Table \ref{Overhead} illustrates some examples.

\end{document}